\documentclass[journal]{IEEEtran} % this one for double.pdf
\usepackage{amsfonts,amssymb,amsmath,amsxtra,amsthm}
\usepackage{array}
\usepackage[utf8]{inputenc} % para os acentos iso
\usepackage{color, graphicx}
\usepackage{lpic}
\usepackage{setspace}

\newcommand{\ie}{i.e.~}

\newcommand{\figrref}[1]{{ Fig.~\ref{#1}}}

\newcommand{\secref}[1]{Section~\ref{#1}}

\newcommand{\comment}[1]{}

\theoremstyle{definition}
\newtheorem{definition}{Definition}

\global\long\def\size{\mathcal{S}}

\usepackage{acronym}
\acrodef{BER}[BER]{bit error rate}
\acrodef{BSC}[BSC]{binary symmetric channel}
\acrodef{DMC}[DMC]{discrete memoryless channel}
\acrodef{LDPC}[LDPC]{low-density parity-check}

\begin{document}
%\title{Random Shuffling with Systematic Coding for Secrecy -- How Large and Degraded Key?}
\title{Analysis of Short Blocklength Codes for Secrecy}

\author{\IEEEauthorblockN{Willie K. Harrison\IEEEauthorrefmark{1}, Dinis Sarmento\IEEEauthorrefmark{3}
Jo\~{a}o P. Vilela\IEEEauthorrefmark{4}, and
Marco Gomes\IEEEauthorrefmark{3} \\
\thanks{
This work was partially funded by the iCIS project under grant CENTRO-07-ST24-FEDER-002003, and the project UID/EEA/50008/2013 - C00356 (WINCE - Wireless Interference and Coding for Secrecy) funded by the Funda\c{c}\~{a}o para a Ci\^{e}ncia e Tecnologia (Portuguese Foundation for Science and
Technology) and conducted at Instituto de Telecomunica\c{c}\~{o}es.
}}

\IEEEauthorblockA{\IEEEauthorrefmark{1}Department of Electrical and Computer Engineering \\
University of Colorado Colorado Springs, Colorado Springs, CO, 80923 \\ Email: wharriso@uccs.edu \\} 
\IEEEauthorblockA{\IEEEauthorrefmark{3}Instituto de Telecomunica\c{c}\~{o}es, Department of Electrical and Computer Engineering\\ University of Coimbra, Coimbra, Portugal \\ Email: dinis.pereira@student.uc.pt, marco@co.it.pt \\}
\IEEEauthorblockA{\IEEEauthorrefmark{4}CISUC and Department of Informatics Engineering, University of Coimbra, Coimbra, Portugal\\ Email: jpvilela@dei.uc.pt} 
}

%\author{
%\thanks{TODO - affiliation}% <-this % stops a space
%}
%\thanks{Manuscript received April 19, 2005; revised December 27, 2012.}}


%\markboth{Journal of \LaTeX\ Class Files,~Vol.~11, No.~4, December~2012}%
%{Shell \MakeLowercase{\textit{et al.}}: Bare Demo of IEEEtran.cls for Journals}

\maketitle

\begin{abstract}
In this paper we provide secrecy metrics applicable to physical-layer coding techniques with finite blocklengths over Gaussian and fading wiretap channel models. Our metrics go beyond some of the known practical secrecy measures, such as bit error rate and security gap, so as to make lower bound probabilistic guarantees on error rates over short blocklengths both preceding and following a secrecy decoder. Our techniques are especially useful in cases where application of traditional information-theoretic security measures is either impractical or simply not yet understood. The metrics can aid both practical system analysis, and practical system design for physical-layer security codes. Furthermore, these new measures fill a void in the current landscape of practical security measures for physical-layer security coding, and may assist in the wide-scale adoption of physical-layer techniques for security in real-world systems. We also show how the new metrics provide techniques for reducing realistic channel models to simpler discrete memoryless wiretap channel equivalents over which existing secrecy code designs may achieve information-theoretic security.
\end{abstract}

%\begin{IEEEkeywords}
%physical-layer security; 
%\end{IEEEkeywords}

 %\ifCLASSOPTIONpeerreview
 \begin{center} \bfseries EDICS Category: COM-OTHS, INF-SECC \end{center}
 %\fi
\IEEEpeerreviewmaketitle

\ifCLASSOPTIONonecolumn
\doublespacing
\fi

\section{Introduction}
\label{sec:intro}
Physical-layer security has attracted much attention of late as a means to provide a keyless layer of security using error-control coding and other physical-layer techniques such as intentional jamming~\cite{Harrison2013,Mukherjee2014}. While traditional information-theoretic secrecy measures have been the preferred vehicles for proving the worth of physical-layer security coding schemes, some channel models remain elusive to this type of analysis~\cite{Ling2014}. In this paper, we provide two new security metrics that apply when blocklengths are finite (and especially when they are short), and when channel models are more representative of real-world environments. 

Coding techniques exist that can achieve strong secrecy, and even semantic secrecy over the binary erasure wiretap channel~\cite{BlochNew}, but in the face of fading, jamming, and otherwise Gaussian noise, there remains a dearth of useful secrecy metrics beyond simple bit-error rates (BER). The one exception is the security gap~\cite{klinc-ldpc_wiretap}, which provides a measure on the required signal-to-noise ratio (SNR) advantage over an eavesdropper to operate at acceptable error rates for friendly parties with an acceptable amount of security over illegitimate receivers. Our metrics go beyond security gap, so as to identify operable regions of SNR for which bit-error rates, even over a short number of bits, are guaranteed to be near 0.5. The basic premise of our techniques is to evaluate \emph{the distribution} of error rates over a small number of bits, such as might be transmitted over a single packet, or within a single coded word, and to make guarantees not only on the mean of the distribution, but rather on, e.g., the 10th percentile or even the 1st percentile of the distribution. A proper tool that allows us to make these claims is the simple cumulative distribution function (CDF) of the error rate over short blocklengths. As one considers percentiles closer to zero, the guarantees of our secrecy metrics are such that every small block of transmitted data either fails to be decoded (for the first metric), or achieves decoder output bit-error rates greater than $0.5-\delta$ (for the second metric). These metrics fill a void in the current landscape of security measures for secrecy codes, and find immediate application in real-world environments.

\begin{figure*}
  \centering
  \ifCLASSOPTIONonecolumn
    \includegraphics[width=\columnwidth]{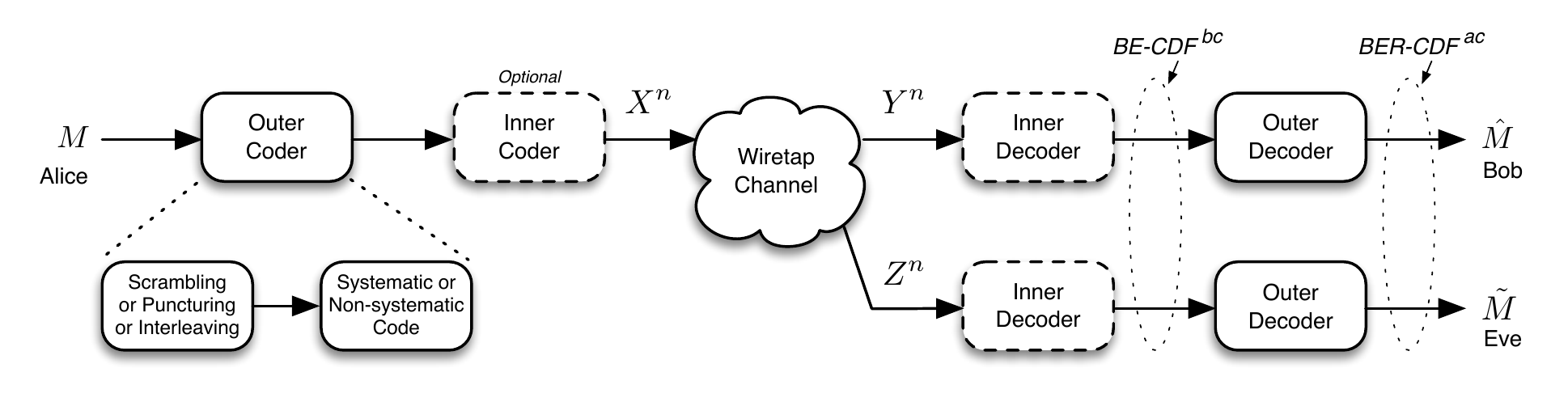}
  \else
    \includegraphics[width=2\columnwidth]{system_setup}
  \fi
  \centering
  \caption{Wiretap channel model assuming a concatenated coding scheme, where the outer code is for secrecy and the inner code is for reliability. Note that the inner code is marked as optional, and if it is removed, then this model reduces to the traditional wiretap channel model. The new metrics presented in this work are BE-CDF$^{bc}$ (where \emph{bc} indicates \emph{before code}), and BER-CDF$^{ac}$ (where \emph{ac} indicates \emph{after code}).}
  \label{fig:system_setup}
\end{figure*}

Consider the wiretap setup as depicted in Fig.~\ref{fig:system_setup}, where the receiver chains for both a legitimate receiver Bob and an eavesdropper Eve are pictured. We consider here a possibly concatenated coding system, where the outer code is for security (and may consist of any number of coding operations as indicated), and the inner code is for reliability. Based on early work over the wiretap channel~\cite{wyner:75,csiszar:78}, we know that there exists a supremum of achievable rates such that both security and reliability can be attained. This rate is called the \emph{secrecy capacity} $C_s$. Unfortunately, the grand majority of all currently known explicit secrecy codes do not provide both reliability and security, but rather offer security as long as the legitimate receiver's channel is noiseless. Explicit code constructions that are exceptions to this rule require that the eavesdropper's channel is degraded from the main legitimate receiver's channel, and only work for discrete memoryless channels~\cite{Harrison2013}. One possible framework for extending these results is to employ a concatenated coding scheme as we illustrate in Fig.~\ref{fig:system_setup}. It should be noted that the inner code in this figure is marked as \emph{optional}, and if it is removed, then the model reduces to the traditional wiretap channel model~\cite{wyner:75}. Thus, although we are considering our new metrics in cases where concatenated codes are used, they remain applicable to the general wiretap case. We note the transmitter Alice encodes a message through all stages of the encoder to produce a length-$n$ codeword $X^n$, which is transmitted over the wiretap channel. Bob and Eve observe their respective signals $Y^n$ and $Z^n$, and both attempt to decode the message, perhaps producing respective message estimates $\hat{M}$ and $\tilde{M}$.

\subsection{An Example}

As a simple example, consider the case where the outer code is just a scrambler, implemented by multiplying the binary length-$k$ message $M$ by a $k\times k$ binary matrix that is invertible in $GF(2)$ at the encoder and its inverse at the decoder. Let's assume that the inner code is a $t$-error correcting code, such as a BCH code. If the channel is a Gaussian or a fading channel, then an information-theoretic security analysis may prove difficult. The alternative is to simulate the concatenated coding scheme at the decoder so as to obtain some guarantee on BER. When this is done, simulations are typically averaged over thousands of runs to obtain an average BER, and although the analysis is simulation driven, the results still only hold asymptotically as blocklengths become very large, just as in an information-theoretic analysis (if it's even possible). 

\begin{figure}
  \centering
  \ifCLASSOPTIONonecolumn
    \includegraphics[width=0.5\columnwidth]{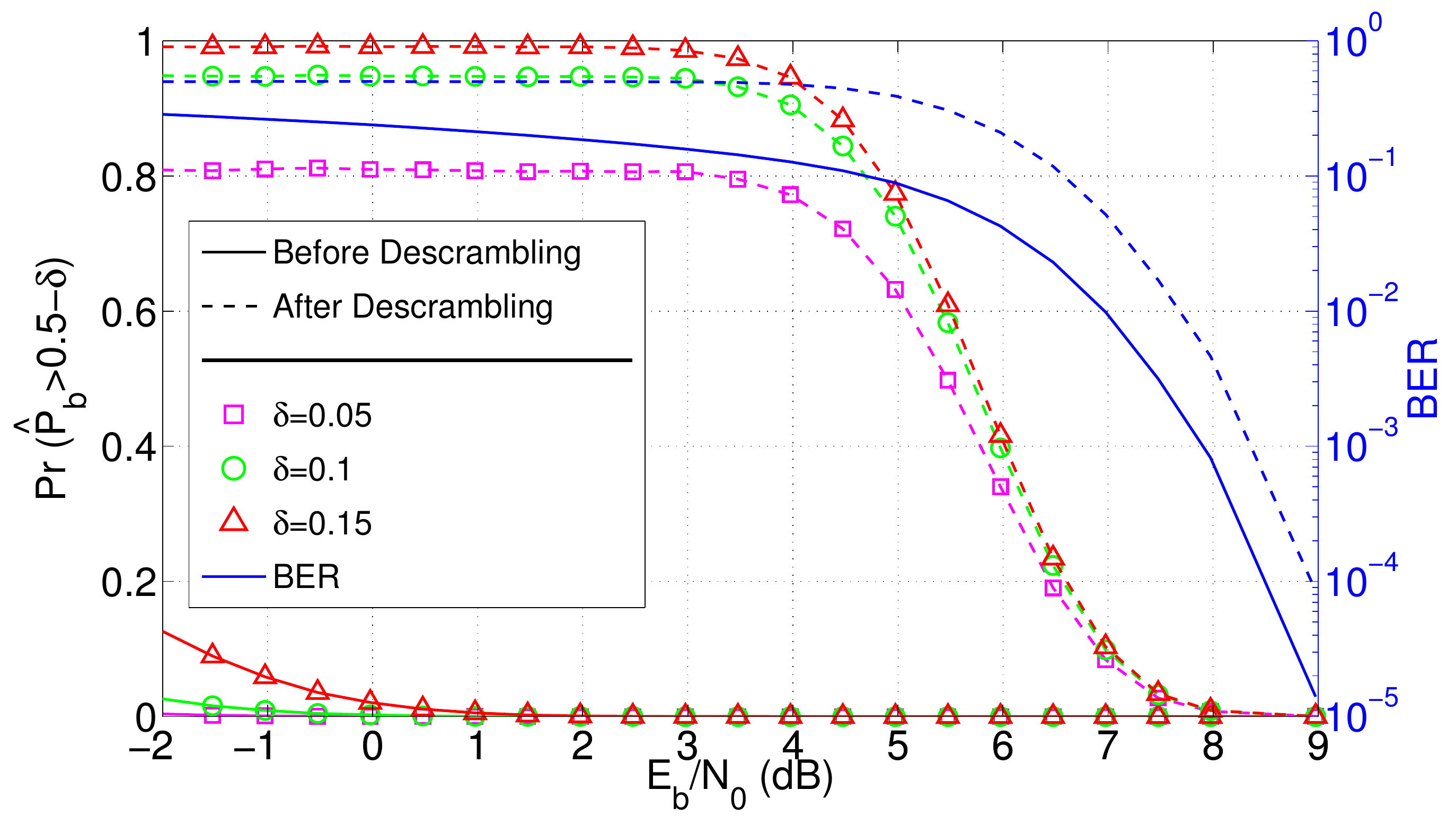}
  \else
    \includegraphics[width=\columnwidth]{Metric_2_scrambling_ebno.pdf}
  \fi
  \centering
  \caption{New security metrics for a simple system where the outer code is a scrambler and the inner code is a BCH($127, 64$) code. BER curves are given in blue with no markers, and $\Pr(\hat{P}_b>0.5-\delta)$ curves are given with markers as indicated to identify the values of $\delta$. Solid lines indicate the location is before the outer coder, while dashed lines indicate the location is after the outer coder.}
  \label{fig:scrambling_metric2}
\end{figure}

We wish to provide probabilistic guarantees of decoder failure and guarantees of low statistical dependence between the message $M$ and an eavesdropper's decoder output message $\tilde{M}$. Despite the fact that BER has several shortcomings as a security metric, it can still be used effectively to estimate decoder outputs when the eavesdropper's attack strategy is known. Our metrics strengthen this approach by considering the entire distribution of possible error rates. In Fig.~\ref{fig:scrambling_metric2} we show the BER both before and after the scrambler in a receiver, and as expected the descrambling operation propagates errors into the message estimate. However, if we'd like to guarantee error rates close to 0.5 in all $k$-bit message estimates at the eavesdropper, it is necessary to consider the entire distribution of error rates over a blocklength of data. We see curves for $\Pr(\hat{P}_b > 0.5-\delta)$ in the figure, where $\hat{P}_b$ can be used to model the proportion of bits in error over one block of $k$ bits either at the input or at the output of the outer decoder, and is a point estimator of the true bit error rate $P_b$. To be more specific, let $B$ be a random variable that represents the number of bits in error over $k$ bits either at the input or the output of the outer decoder. Then
\begin{equation}
 \hat{P}_b = \frac{B}{k},
\end{equation}
and is coincidentally the maximum likelihood estimator for the bit error rate $P_b$ given $k$ independent observations~\cite{ProbAndStats}. While the errors in $k$ received bits comprising a single transmitted codeword are likely not independent at the output of a decoder, we will address this concern later in Section~\ref{sec:metric2}.
Notice in Fig.~\ref{fig:scrambling_metric2} that if we want $\Pr(\hat{P}_b > 0.5-\delta)$ after the decoder to get close to one, then we need to allow $\delta > 0.15$ for this scheme, and somehow ensure that Eve's $E_b/N_0$ is no better than 3 dB. Also note that we use this simple example to showcase the general applicability of the new metrics, as comparing error rates before and after the outer decoder gives one method for quantifying the contribution of the descrambler to the data received by the eavesdropper, but we are not proposing scrambling with BCH codes as a security solution.

\subsection{Outline}

Throughout this paper, we will let SNR designate the signal-to-noise ratio as measured by the channel, meaning the energy per transmitted bit over the noise power spectral density $N_0$. $E_b/N_0$ will be the energy per information bit divided by $N_0$. The two are related by the overall rate $R$ of the concatenated coding scheme so that $\text{SNR} = RE_b/N_0$ for BPSK transmission.

The rest of the paper is organized as follows. First, we survey the landscape of secrecy metrics for physical-layer security coding schemes in Section~\ref{sec:metrics}. We then point out some shortcomings and motivate the need for additional practical metrics in Section~\ref{sec:short}. Since the main contribution of this paper is the introduction of new secrecy metrics, these two sections are absolutely crucial. In Section~\ref{sec:short} we also highlight the cases for which our metrics are superior to both information-theoretic and BER-based existing metrics, and point out their limitations. Sections~\ref{sec:metric1} and~\ref{sec:metric2} provide our new metrics BE-CDF$^{bc}$ and BER-CDF$^{ac}$, respectively, with definitions and clarifying examples. Finally, we show a use case of these metrics in a more complicated concatenated coding scheme in Section~\ref{sec:apps}, and indicate how the scheme may be used directly for secrecy, or used to provide a discrete memoryless wiretap channel equivalent over which additional secrecy codes may be used to achieve information-theoretic security. We offer some comments by way of conclusion in Section~\ref{sec:conclusion}. 

%\todo{brief overview of metrics; nice motivating example of the need for a new one, need to introduce the wiretap channel here also with signal definitions $M$, $X^n$, $Y^n$, and $Z^n$}

%\jp{Food for thought (29th December):
%  \begin{itemize}
%  %\item consider a concatenated coding scheme such as presented in figure 1, where an inner code (e.g.~LDPC or turbo code) is used for reliability, while an outer code provides a source for secrecy (e.g.~scrambling, puncturing, or interleaving + jamming)
%  \item two metrics: BE-CCDF$^{bc}$ vs BER-CCDF$^{ac}$ (strong BER? caution on calling this strong because this is still a rate...)
%  \item bc -- before the outer code; ac -- after the outer code (\ie at the exit of the overall coding scheme)
%  \item the use of these metrics enables the analysis and development of coding schemes where an inner code is selected for reliability and an outer code provides a source of randomness/advantage that is used in favor of confidentiality
%  \item in our random interleaving for secrecy scheme the use of $t$-error correcting codes warants a high BE-CCDF$^{bc}$ but does not necessarily provide a high BER at the exit of the inner LDPC code; however, due to the binary decoding nature of the $t$-error correcting codes, our scheme provides a high/tight BER-CCDF$^{ac}$ (very close to 0.5) at the exit of the coding for secrecy scheme, \ie after the outer code
%  \end{itemize}
%}

\section{Secrecy Metrics}
\label{sec:metrics}
The secrecy metric space has progressively become more dense, particularly over the last few decades. The initial secrecy coding metric posed by Shannon in the late 1940's was that of \emph{perfect secrecy}~\cite{shannon49secrecy}. A code is said to achieve perfect secrecy if 
\begin{equation}
  I(M;X^n) = 0,
\end{equation} 
or, alternatively, if the equivocation $H(M|X^n)$ is equal to the entropy of the message $H(M)$. Perfect secrecy indicates that the coded message tells you nothing about the message itself. Shannon introduced the notion through the coding scheme of the one-time pad, and promptly proved that it was impossible to achieve perfect secrecy in a scheme where the entropy of a secret key is not at least as much as the entropy of the message itself, making the notion completely impractical.

In the mid-seventies, Wyner~\cite{wyner:75} introduced an additional metric for secrecy that is known today as \emph{weak secrecy}. A scheme is said to achieve weak secrecy if 
\begin{equation}
  \lim_{n\rightarrow\infty} \frac{1}{n}I(M;Z^n) = 0.
\end{equation}
This metric introduced the idea of coding for secrecy in earnest because the results indicated that it was actually possible to achieve weak secrecy in a practical system. After all, this criterion does not require that the coded message $X^n$ leaks no information about $M$, but rather that the eavesdropper's observation $Z^n$ must leak a sufficiently small amount of information about $M$ such that the $1/n$ factor can still drive the quantity to zero. With this new notion of secrecy, came the idea of secrecy capacity $C_s$ which was originally defined as the supremum of coding rates that can achieve weak secrecy against a passive eavesdropper as a function of the wiretap channel parameters, while maintaining arbitrarily low probability of decoding error at the legitimate receiver. As long as the legitimate parties are able to leverage an advantage over the eavesdropper so that the effective main channel is \emph{less noisy}~\cite{csiszar:78} than the eavesdropper's channel, then $C_s > 0$, which indicates that private communications are theoretically possible.

Weak secrecy was shown to be insufficient in many cases~\cite{blochbarros-phylayersec}, and Maurer later defined a stronger metric known as \emph{strong secrecy}~\cite{Maurer1994}, where a scheme is said to achieve strong secrecy if 
\begin{equation}
  \lim_{n\rightarrow\infty} I(M;Z^n) = 0.
\end{equation}

It was recently noted in \cite{Bellare2012} that even strong secrecy may not be sufficient for some applications because the assumption is often made that message symbols are random and uniformly distributed over the message alphabet. Of course, in cryptographic scenarios, the messages are never perfectly random and uniform, and it is known that in practice there really is no universal compression algorithm that can provide such messages at the input of secrecy encoders. Thus, we have an even stronger notion of secrecy called \emph{mutual information security} which is achieved if 
\begin{equation}
  \lim_{n\rightarrow\infty} \max\limits_{p_M}\{I(M;Z^n)\} = 0.
\end{equation}
Here we maximize $I(M;Z^n)$ over all possible message distributions $p_M$. It is also shown in~\cite{Bellare2012} that this notion of secrecy is equivalent to \emph{distinguishing security} and \emph{semantic security}.

%introduced a stronger notion of security still for the wiretap channel known as semantic security, which is inspired by cryptographic security principles, but is also information-theoretic. The idea here is that we define the advantage between $M$ and $Z^n$ as}
%\begin{align}
%  \textrm{Adv}(M; Z^n) & = \nonumber \\  \max\limits_{f,p_M} \left(\sum\limits_{z^n\in\mathcal{Z}^n} \right. & p_{Z^n}(z^n) \max\limits_{f_i\in\mathrm{supp}(f)} \Pr(f(M) = f_i|z^n) - \nonumber \\  \max\limits_{f_i\in\mathrm{supp}(f)} \Pr(f(M)& \left. =f_i)\right), 
%\end{align} 
%\textcolor{green}{so that $\textrm{Adv}(M; Z^n) < \epsilon$, indicates that observing $z^n$ increases the probability of guessing some function of $M$ by no more than $\epsilon$~\cite{BlochNew}. Notice this is calculated over all possible distributions on the message $p_M$, which allows us to think of nonrandom messages such as we would expect in any meaningful transmitted message.

Although it took over 30 years after Wyner introduced weak secrecy for an explicit code design to emerge that could achieve it~\cite{ThangarajDCMM07}, it has already been shown that codes exist that can achieve both strong and semantic secrecy, albeit over simple wiretap channel models~\cite{BlochNew,Bellare2012}, and surprisingly, the secrecy capacity defined using strong or semantic security is provably the same as that defined by the weak secrecy metric~\cite{Maurer2000,BlochNew}. 

Although this list of information-theoretic measures is impressive, there remain several wiretap channel models that have proved elusive to explicit code designs where information-theoretic security can be guaranteed. Thus, over channels that are more representative of real world communications, such as the Gaussian wiretap channel or fading channel scenarios, there have been additional security metrics developed. For example, the authors in~\cite{klinc-ldpc_wiretap,baldi2012coding} used bit-error rate (BER) at the output of a decoder as a more practical means of security measure. This metric can be simulated in a straight-forward manner, just as is done for traditional error-correcting codes. The authors in~\cite{klinc-ldpc_wiretap} developed a new secrecy metric by identifying a target BER for the legitimate receiver, as well as a target BER for an eavesdropper, and found the SNRs that would achieve each of these targets. The \emph{security gap} was then defined as the difference between these two SNR values in dB (or a ratio of the two linear values). The security gap tells a designer what the required advantage is for obtaining the desired security and reliability performance, and threshold operating points for achieving both. 

Authors in~\cite{Harrison2011} studied coding mechanisms that provided \emph{degrees of freedom} in an eavesdropper's decoder output, where no information about certain bits could be obtained, forcing an attacker to guess the bits associated with the degrees of freedom in the decoder. This notion was similar to an information-theoretic security approach in the sense that the information could not be attained through any degree of processing, but was also very much unlike an information-theoretic security approach because it restricted an attacker to a specific attack strategy. 

%\todo{Still to consider?}
%
%\begin{itemize}
%%\item Mutual information \cite{shannon49secrecy}
%%\item Secrecy capacity \cite{wyner:75}
%% \item Ergodic secrecy capacity \cite{}
%\item Probability of secrecy outage / probability of secrecy capacity outage \cite{secrecy_capacity,BlochBRM08}
%%\item Equivocation (perfect, weak and strong secrecy) \cite{}
%%\item Bit error rate \cite{demijan-ldpc09,klinc-ldpc_wiretap}
%%\item Security gap \cite{demijan-ldpc09,klinc-ldpc_wiretap}
%%\item Secure degrees of freedom \cite{}
%%\item Semantic security \cite{}
%\end{itemize}

\section{Shortcomings of Current Security Metrics}
\label{sec:short}

The metrics of the previous section give many techniques for analyzing the security achieved by specific coding schemes.  Developing wiretap codes that are able to reach the envisioned secrecy capacity for more practical channel models remains a formidable challenge, and performing the information-theoretic analysis is oftentimes deemed intractable. The information-theoretic measures are still the most desirable where possible to apply, but they also have another weakness in the sense that they lead to codes that are designed to meet a secrecy criterion in an asymptotic blocklength regime only, thus limiting their applicability in real systems that require short blocklength codes.

On the other hand, one should be careful when performing security analysis that relies only on BER-based measures, because high error rates do not necessarily indicate that some information has not been leaked. In fact, modern cryptography is based on computational security that does leak the information about the message. These systems work not because of an information-theoretic guarantee, but rather due to there being no known computationally efficient algorithm that can find the solution in any reasonable amount of time with any realistic amount of computing power unless the key is known. Thus we see that despite not achieving an information-theoretic security measure, cryptosystems remain useful because they attain security in a more practical/applied sense. In a similar way, BER security analysis assumes the best known decoder/attack, and makes calculations assuming an eavesdropper uses that attack. While BER may provide some useful information about the quality of the received data or the decoder output at the eavesdropper, BER calculations are still made by averaging large amounts of data, and are therefore only reliable as blocklengths get large. 

The metrics we introduce over the next two sections of this paper take a BER approach, but rather than calculating simple averages, make use of our knowledge of the CDF of bit error rates over small blocks of data to provide lower bounds on error rates through the receiver decoder chain. Making this fundamental change in how BER is used to analyze security in a system, allows us to make stronger guarantees about the performance of secrecy codes in the short blocklength regime. This is something that none of the metrics in Section~\ref{sec:metrics} can provide due to the way the analysis is completed either as blocklength goes to infinity, or as simulations are averaged over thousands of independent runs. Using the new metrics, we also maintain the ease of simulation-based characterization of security (which is particularly helpful when realistic channel models are considered, where it is not known how to provide an information-theoretic analysis). Table \ref{tab:summary} outlines the utility of each currently known physical-layer security metric, and indicates the contribution of our new metrics lies in ease of computation and providing the strongest guarantee yet for analyzing finite blocklength code designs.
%, but do not require us to rely on averages over thousands of runs to analyze short-term performance in a system. It is exactly these characteristics in a security metric that can allow designers to thoroughly analyze secrecy codes and attain security guarantees without the need for difficult theoretical analysis with each new coding setup. Such metrics can make secrecy codes practical enough for implementation in real-world communications systems.

%To address these issues, recent works \cite{klinc-ldpc_wiretap,baldi2012coding} have focused on a more operational security metric -- the bit-error-rate (BER), in that Bob should be able to achieve low enough BER levels for reliability while Eve is confronted with a BER close to 0.5. However, with the BER being an average metric, we may have at times fewer than the expected number of errors at Eve, which may be unacceptable from a security perspective.

%\note{(JP) I've added this text here because I needed a bit of context -- feel free to change!}

\begin{table*}[]
\caption{Summary of current physical-layer security metrics, highlighting some of their pros and cons. Here we see that although our new metrics cannot provide information-theoretic security, they are best in class among the error-rate secrecy metrics. Note: w.p. means \emph{with probability}.}
\begin{center}
\ifCLASSOPTIONonecolumn
  \scriptsize
\else
  \footnotesize
\fi
\begin{tabular}{c|c||c|c|c|c|c}
  & & Directly applicable & Easily computable & Information-theoretic & Strongest & Achievable \\  
  Class & Metric & to short codes & in general & secrecy guarantees & in class & in practice \\\hline \hline
 Impractical & Perfect Secrecy & Yes & No & Yes & Yes & No \\ \hline
 Info-theoretic & Weak Secrecy & No & No & Yes & No & Yes \\
 Info-theoretic & Strong Secrecy & No & No & Yes & No & Yes \\
 Info-theoretic & Semantic Secrecy & No & No & Yes & Yes & Yes \\ \hline
 Error rate & BER & No & Yes & No & No & Yes (BER $\approx 0.5$) \\
 Error rate & Security gap & No & Yes & No & No & Yes (security gap $< 0$ dB) \\
 Error rate & BE-CDF$^{bc}$ & Yes & Yes & No & No & Yes (decoder failure w. p. $\approx 1$) \\
 Error rate & BER-CDF$^{ac}$ & Yes & Yes & No & Yes & Yes (high error rates w. p. $\approx 1$) \\ 
 \hline
 
\end{tabular}
\end{center}
\label{tab:summary}
\end{table*}%

\section{The Bit Error-Cumulative Distribution Function}
\label{sec:metric1}

\begin{figure}
  \centering
   \ifCLASSOPTIONonecolumn
    \includegraphics[width=0.5\columnwidth]{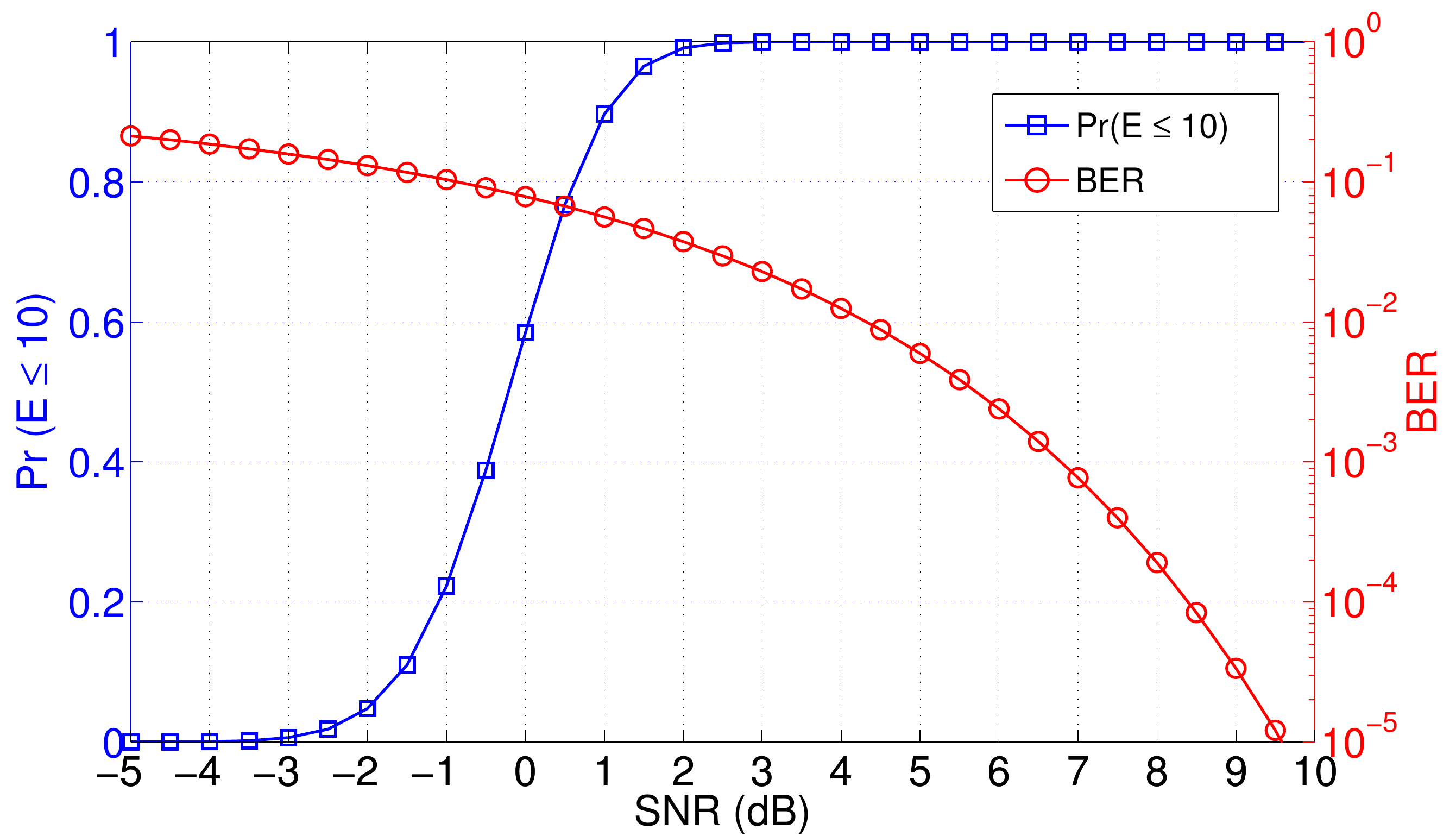}
   \else
    \includegraphics[width=\columnwidth]{ber_becdf-awgn.pdf}
   \fi
  \centering
  \caption{Bit error probability and probability of having fewer than or equal to $10$ errors for an AWGN channel with BPSK modulation.}
  \label{fig:ber_becdf-awgn}
\end{figure}

%While the BER avows a metric that can be used in practical scenarios with codes of finite blocklength, it can be misleading in designing coding for secrecy schemes. 
Let us consider an AWGN channel with BPSK modulation, for which the BER (depicted in \figrref{fig:ber_becdf-awgn}) is given by \cite{MoonCoding}
\begin{equation}
P_b = \frac{1}{2}\mathrm{erfc}\left(\sqrt{\text{SNR}}\right).
\label{Pb}
\end{equation}

A $t$-error correcting code of length 127 that is able to correct up to 10 errors can recover from a BER of $\frac{10}{127}\approx 0.079$ assuming uniform error distribution, but errors over short blocks of data are not guaranteed to occur so uniformly. Let $E$ be the number of bit errors in a block of $n$ bits. For a transmitted word of size $n$ with independent bit errors, the probability of having fewer than or equal to $t$ errors, $\Pr(E\leq t)$ can be straightforwardly obtained from (\ref{Pb}) as
\begin{equation}
\Pr(E\leq t) = \sum_{i=0}^t {n \choose i} {P_b}^i (1-P_b)^{n-i}.
\end{equation}
Let us now consider two operating points of \figrref{fig:ber_becdf-awgn}: (a) $\text{SNR}=0$ dB that leads to a BER close to the $0.079$ that the code supports, and (b) $\text{SNR}=-3$ dB, that leads to a BER $\approx 0.16$. Looking at $\Pr(E\leq 10)$ in the same figure, for $\text{SNR}=0$ dB we have $\Pr(E\leq 10) \approx 0.58$, meaning that the code would still succeed more than half of the time. For $\text{SNR}=-3$ dB, we get $\Pr(E\leq 10) \approx 0.006$, which indicates that the decoder will fail over 99\% of the time, yet with a BER far from $0.5$. Also note that the curve for $\Pr(E\leq 10)$ approaches zero for low SNR values, with the BER still far from the idealized $0.5$ value. With this in mind, the question arises of how close to BER$=0.5$ is close enough for security purposes?

 %% \jp{While the BER avows a metric that can be used in practical scenarios with codes of finite blocklength, it suffers from the well-known disadvantages of average-based metrics, which turn out inappropriate for security analysis of coding schemes. For example, a $t$-error correcting code of length 127 code that is able to correct up to 10 errors is able to recover from a BER of $\frac{10}{127}\approx 0.079$. Considering an AWGN channel with BPSK, this BER level requires a $E_b/N_0\approx 0$dB. Numerical results show that under this setup the probability of having more than $10$ errors is $\approx 0.42$, meaning that the code would still succeed more than half of the time, which is unacceptable from a security perspective.

To address this issue, we look to the distribution of errors of transmitted data and propose the first of two new secrecy metrics. 
\begin{definition}[Bit Error Cumulative Distribution Function]
The bit error cumulative distribution function, BE-CDF$^{bc}$($t$,~SNR,~$\size_m$,~$\mathcal{C}_i$), gives us the probability of having $t$ or less errors, $\Pr(E\leq t)$, as a function of the SNR for a message of size $\size_m$, encoded with a code $\mathcal{C}_i$ (refers to the optional \emph{inner} code).
%\note{as a function of the SNR to make it more general -- in our random interleaving4secrecy paper, this was a function of the jamming power, which could be translated to a reduction of SNR; changed it to $E < t$ so that it grows with the SNR}
%% \todo{make definition more generic -- it can be a function of the SNR or something else (in our paper it was the jamming power); also it can be used without any code, to analyze errors right after the channel}
\end{definition}
From this metric we can deduce the probability of having more than $t$ errors in a block of data, giving us the power to predict the likelihood of decoder failure when the code is a $t$-error correcting code such as a BCH code. This information is useful for identifying acceptable SNR operating points for both friendly parties and eavesdroppers~\cite{vilela2015}. Notice from Fig.~\ref{fig:system_setup} that we measure this metric \emph{before} the outer \emph{code} (hence the superscript \emph{bc}) in a concatenated coding scheme, i.e. prior to the secrecy code. Because of this, we choose to use SNR, rather than $E_b/N_0$ to show the results, although the conversion can be made if desired.
%Our goal with this metric is to assess the reliability and confidentiality levels for communication with $t$-error correction codes, in connection with aspects such as the number $t$ of acceptable errors, the operating SNR and the size of the message.

\subsection{Analysis}
This metric can also be used to fine tune the security and reliability levels of a coding scheme that relies on $t$-error correcting codes. For example, if we assume no inner code and set the outer code to a BCH($127, 64$) code that corrects up to 10 errors, and if we want a reliability level of $\Pr(E\leq 10)>0.99$, Bob would have to operate at an SNR above $1.95$ dB as indicated in Fig.~\ref{fig:ber_becdf-awgn}. For a confidentiality level of $0.99$, \ie $\Pr(E\leq 10)<0.01$, Eve would need to operate at SNR below $-2.78$ dB.

While relevant reliability and confidentiality levels with a reasonable SNR gap between Bob and Eve may seem illusive with simple coding schemes such as the mentioned BCH code, this metric enables the selection of $t$-error correcting codes that can be used in more evolved concatenated coding schemes combined with the generation of interference \cite{vilela2015} to provide desired levels of reliability and confidentiality, as will be described in \secref{sec:apps}.

%% \subsection{Use Cases \note{to go into section VI}}
%% \todo{brief description of random interleaving for secrecy coding scheme \cite{vilela2015}}

\section{The Bit Error Rate-Cumulative Distribution Function}
\label{sec:metric2}

The BE-CDF$^{bc}$ allows us to guarantee failed decoding with high probability over certain SNR regions for $t$-error correcting codes. However, a failed decoder does not necessarily imply that the eavesdropper cannot obtain most of the message bits at the output. Hence, in this section we introduce a metric that can guarantee decoder failure with BER close to 0.5 in the estimated message bits to strengthen the security guarantee. For this section, let $\hat{P}_b$ be the measured proportion of bit errors at the output of an error-correcting decoder measured over $\size_b$ decoded message bits. For the case where the code being used is a block ($n, k$) code, it makes sense to let $\size_b$ be an integer multiple of $k$. The metric we propose in this section allows a user to specify a required error rate at the output of the eavesdropper's error-control decoder over $\size_b$ bits using the probability that $\hat{P}_b > 0.5-\delta$ for any $\delta$ desired.
\begin{definition}[Bit Error Rate-Cumulative Distribution Function]
The Bit Error Rate-Cumulative Distribution Function, BER-CDF$^{ac}$($\delta$, $E_b/N_0$, $\size_b$, $\mathcal{C}$) is the quantity
\begin{equation}
  \Pr(\hat{P}_b > 0.5-\delta)
\end{equation}
calculated over $\size_b$ estimated message bits for a code $\mathcal{C}$ as a function of $E_b/N_0$, where $\mathcal{C}$ may be the concatenation of an (optional) inner code $\mathcal{C}_i$ and an outer code $\mathcal{C}_o$.
\end{definition}
We note that the \emph{ac} exponent indicates that the metric is measured \emph{after the code}. Since the inner code is shown to be optional in Fig.~\ref{fig:system_setup}, this is referring to the outer (secrecy) code. Also, because we are calculating this metric after the decoder, it makes sense to use $E_b/N_0$, rather than SNR. Finally, we should note that this metric is actually the complement to the CDF, but we choose to use a consistent nomenclature to that of the BE-CDF$^{bc}$. These two metrics packaged in a pair provide valuable design information so as to achieve both reliability and secrecy.
\subsection{Analysis}

%\todo{systematic vs non-systematic; if systematic, can go for uncoded when BER too low}

The BER-CDF$^{ac}$ allows us to guarantee decoder failure with high probability in addition to high BER over short blocks of $\size_b$ bits at the output of the decoder. Although the metric is not information-theoretic, it comes much closer to the information-theoretic definitions of secrecy than the BE-CDF$^{bc}$ by limiting the amount of useful information to an eavesdropper (as tends to happen with high BER). That is, for a scheme that guarantees high BER using the BER-CDF$^{ac}$ metric, it is unlikely that the decoder will fail and yet provide small BER at the output. Notice that this metric is also much more robust than simply considering the average BER, and examples are shown in the following section of the paper.  Similarly as with our BE-CDF$^{bc}$ metric, we now ensure that the entire distribution of BER values for a specific length of text $\size_b$ is within an acceptable security region.

Recall that $\hat{P}_b$ is the estimator of the error rate $P_b$ at the output of the final decoder over a short blocklength of $\size_b$ bits. If we assume that each bit at the output of the decoder is in error independently with probability $P_b$, then the random variable $P_n = \size_b\hat{P}_b$ models the number of errors in a block of $\size_b$ bits, and is distributed according to the binomial distribution with parameters $\mu = P_b$, and $\sigma^2 = \size_bP_b(1-P_b)$. This means we can calculate the metric exactly as
\begin{align}
 \Pr(\hat{P}_b > 0.5 - \delta) & = \Pr[P_n > \size_b(0.5-\delta)] \nonumber \\
 & = 1 - \sum\limits_{x=0}^{\lfloor \size_b(0.5-\delta) \rfloor} {\size_b \choose x} P_b^x(1-P_b)^{\size_b-x}. 
\end{align}
Although the exact expression can be derived in this case, the assumption of i.i.d. errors is not likely to hold in practice, $P_b$ may be unknown, and the calculation itself would be time intensive, or require approximation using the Gaussian distribution~\cite{ProbAndStats}. Thus, in practice, it makes more sense to calculate the metric using straightforward Monte Carlo simulations.
%This metric is a function of $\delta$, the desired guaranteed offset from 0.5 BER, the code parameters including blocklength, etc., the number of $l$ message bits considered in each block of data (perhaps an integer multiple of $k$), the modulation scheme, and the signal-to-noise ratio (SNR) at the receiver. 

\begin{figure}
  \centering
   \ifCLASSOPTIONonecolumn
    \includegraphics[width=0.5\columnwidth]{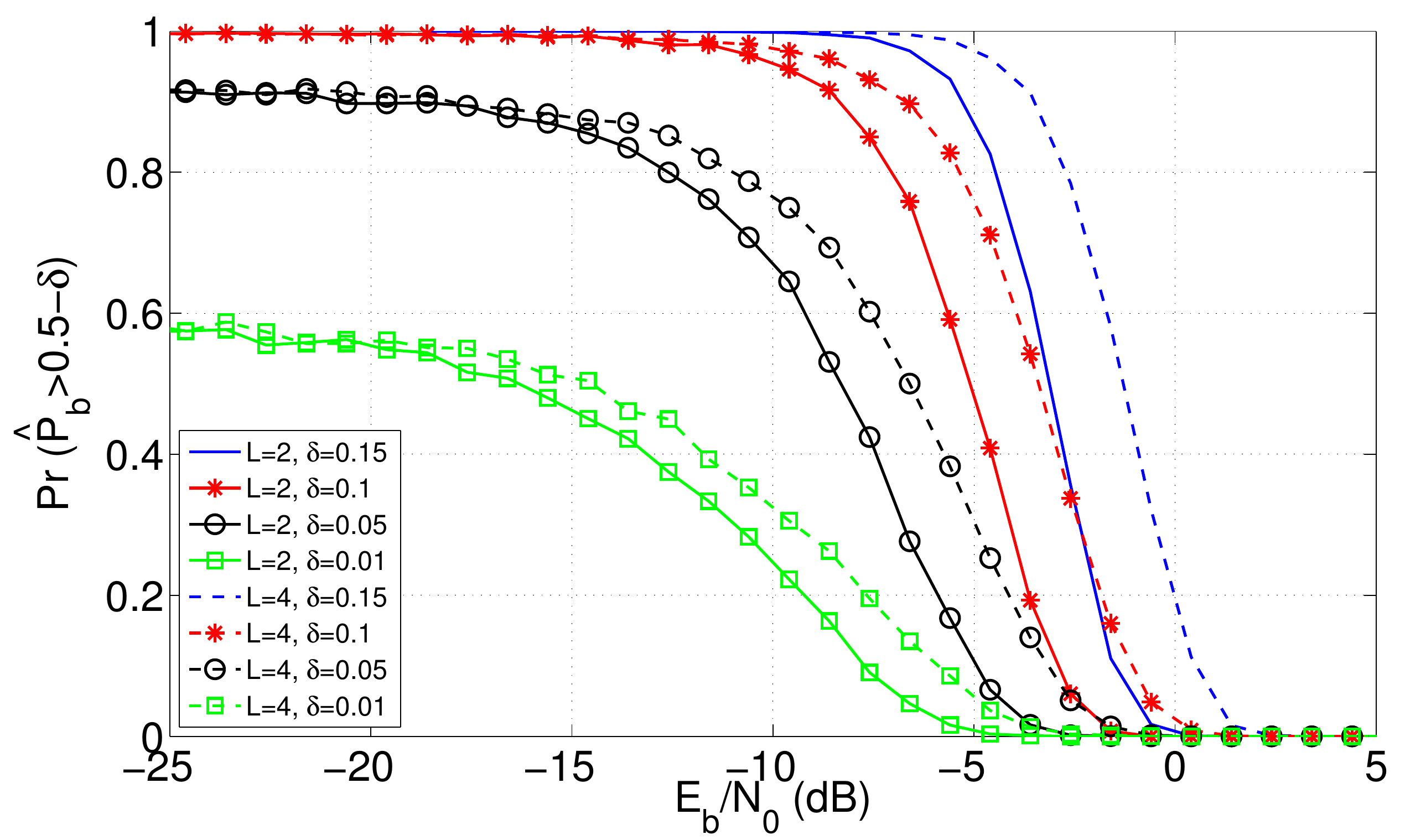}
   \else
    \includegraphics[width=\columnwidth]{fig4_ebno}
   \fi
  \centering
  \caption{Depicts the BER-CDF$^{ac}$ metric $\Pr(\hat{P}_b > 0.5-\delta)$ for the BCH($127, 92$) code for $\size_b = 2 \times 92 = 184$ using $L$-order DPSK modulation. Notice that for some $\delta$ values, the BER-CDF$^{ac}$ approaches one, where other curves appear to be bounded away from one.}
%\vspace{-0.6cm}
  \label{fig:strongBERexample}
\end{figure}

By way of example, consider $\Pr(\hat{P}_b > 0.5-\delta)$ as plotted for a BCH($127, 92$) code as the outer code with several varying sets of parameters as portrayed in Fig.~\ref{fig:strongBERexample}. Each case presented uses $\size_b = 92\times 2 = 184$ so as to allow a $L = 4$ order modulation scheme without zero-padding. The modulation scheme was chosen arbitrarily to be differential phase shift keying (DPSK), and is either binary or quaternary as indicated in the legend. Beyond this, we consider different $\delta$ values as shown. Although there exist $E_b/N_0$ values for which the decoder fails with probability close to one, unless the resultant BER is greater than $(0.5-\delta)$ with high probability, the metric will not approach one in the limit as $E_b/N_0 \rightarrow -\infty$.

Notice that the value the BER-CDF$^{ac}$ approaches as $E_b/N_0 \rightarrow -\infty$ is strongly linked to $\delta$, which makes perfect sense. As $\delta$ grows, it is more possible to fit the entire distribution of BER above the $(0.5-\delta)$ threshold. This observation indicates that for any particular coding scenario, there may in fact exist a minimum $\delta$ for which the BER-CDF$^{ac}$ can be made to go to one as $E_b/N_0 \rightarrow -\infty$. Also notice in Fig.~\ref{fig:strongBERexample} that increasing the order of the digital modulation scheme can bring about an effective shift towards better security. When $\Pr(\hat{P}_b > 0.5-\delta)$ is bounded away from one, we are viewing the random corrective capabilities of the code even when the signal is completely overwhelmed by noise. Certainly, we can do better by increasing $\size_b$ or the dimensions of the code as well, but the utility of this metric is that we can get a clear picture for what happens when $\size_b$ is small, thus providing small blocklength security analysis in practical physical-layer security system designs.

Let us consider the limiting value of the BER-CDF$^{ac}$ as $E_b/N_0 \rightarrow -\infty$. Clearly this quantity is a function of $\delta$ and $\size_b$, and can be calculated by recognizing that $\hat{P}_b$ is a sample mean of Bernoulli random variables $X_i$ where
\begin{equation}
 X_i = \begin{cases} 1 & \mbox{if bit $i$ is in error,} \\ 0 & \mbox{otherwise.} \end{cases}
\end{equation}
Let $\Pr(X_i = 1) = P_b$ as before. Then specifically,
\begin{equation}
  \hat{P}_b = \frac{1}{\size_b}\sum\limits_{i=1}^{\size_b} X_i,
\end{equation}
and by the central limit theorem $\hat{P}_b \sim \mathcal{N}(P_b,\, \frac{P_b(1-P_b)}{\size_b})$. Clearly, this is true in the limit as $\size_b$ gets large, but even for small and moderate blocklength sizes, the central limit theorem still provides a good approximate distribution. 

\begin{figure}
  \centering
   \ifCLASSOPTIONonecolumn
    \includegraphics[width=0.5\columnwidth]{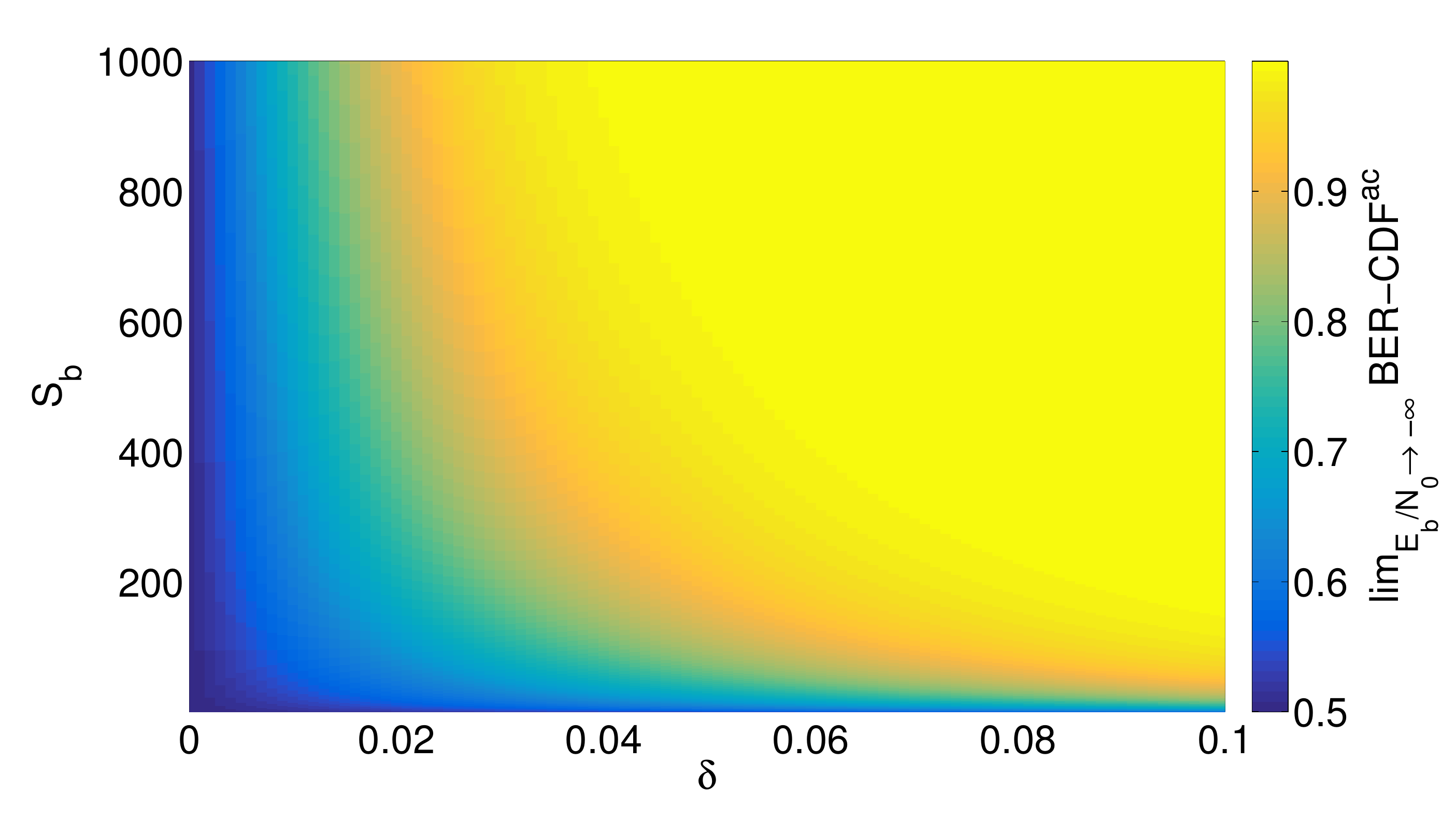}
   \else
    \includegraphics[width=\columnwidth]{limitingMetric2}
   \fi
  \centering
  \caption{Limiting value of the BER-CDF$^{ac}$ metric as $E_b/N_0$ goes to $-\infty$ as a function of $\delta$ and $\size_b$.}
%\vspace{-0.6cm}
  \label{fig:limitMetric2}
\end{figure}

In the limit as $E_b/N_0\rightarrow -\infty$, we also have $P_b\rightarrow 0.5$, and $\hat{P}_b \sim \mathcal{N}(0.5,\frac{0.25}{\size_b})$. Using the classic Gaussian standardization technique~\cite{ProbAndStats}, we find that
\begin{equation}
  \lim\limits_{E_b/N_0\rightarrow -\infty} \Pr(\hat{P}_b > 0.5-\delta) = Q\left(-2\delta\sqrt{\size_b}\right).
\end{equation}
This limiting value of the BER-CDF$^{ac}$ is shown in Fig.~\ref{fig:limitMetric2} over a range of $\delta$ and $\size_b$ values. These results can aid system designers in choosing $\size_b$ (or $k$) in outer codes appropriately so as to supply a desired BER-CDF$^{ac}$. Once $\size_b$ is chosen, we also have a best possible value for the metric over which any coding scheme can be compared. One characteristic of good secrecy codes is that they will transition from zero to the limiting value in this metric over a very short range of $E_b/N_0$.

%\subsection{Use Cases}
%\todo{puncturing \cite{klinc-ldpc_wiretap} and scrambling \cite{baldi2012coding} for secrecy}

\section{Application}
\label{sec:apps}
In this section, we show how the concatenated coding system from~\cite{vilela2015} measures up using the two new metrics, and discuss the utility of the system as a result of its BE-CDF$^{bc}$ and its BER-CDF$^{ac}$ curves. It should also be noted that~\cite{vilela2015} goes through a design process based on the BE-CDF$^{bc}$ for this coding scheme. Although we do briefly outline the scheme and one possible design process here, the interested reader is directed to the original work for further details. Finally, we indicate how our new metrics may be combined with this coding scheme to provide \emph{effective} discrete memoryless wiretap channel equivalents over which other secrecy coding schemes may be implemented to achieve information-theoretic security.

\subsection{System Setup}
\label{sec:scheme}
The system analyzed in this section follows the general concatenated coding framework outlined in Fig.~\ref{fig:system_setup}. The outer code can actually be considered as two encodings, where the message is interleaved according to a secret key $K$ (drawn from the space of possible permutations on $\size_m$ input message bits), and the key is encoded separately from the message using a BCH($127, 64$) code that is capable of correcting 10 errors. The interleaved message and the encoded key are then appended together, and this constitutes the outer code. An LDPC($1056, 880$) code is then used as the inner code, which is applied to the appended message and key to form a codeword suitable for transmission over a noisy channel.  Recall from Fig.~\ref{fig:system_setup} that the general concatenated framework is such that the outer code is intended to achieve the secrecy requirements of the system, while the inner code is used to achieve reliability for Bob. 

In this system, however, there is more at play than just the coding schemes. When the encoded data that are associated with the key $K$ are transmitted over the channel they are intentionally jammed by some friendly network user with jamming power equal to a fraction $\alpha$ of Alice's transmit power. The idea is to give Bob an advantage because of his location or knowledge of the jamming signal so that the jamming affects him only minimally, while an eavesdropper has no information about the jamming signal and/or is positioned in a geographic location that does not afford her the same advantage as Bob~\cite{vilela2015,vilela_positionjamming,Vilela-DSPAN13}. Since the jamming is only applied to the encoded bits associated with the interleaving key, reliability in the system also stems from Bob being able to recover the key for deinterleaving, while security in the system depends on Eve being unable to recover the interleaving key. Data are transmitted over a Gaussian wiretap channel using BPSK modulation.

The receiving decoders at Bob and Eve apply a soft-decoding algorithm for the LDPC code, and the BCH decoder can then correct no more than 10 errors in the key bits. The goal is to reliably keep the errors at the output of the LDPC decoder at no more than 10 for Bob, and above 10 for Eve for each transmitted key block, as the key bits must be used to deinterleave the message bits at the final step of the decoder. The mapping of keys to interleavers is such that any errors in the estimated key result in high error rates in the deinterleaved message, even when the interleaved message bits are recovered exactly~\cite{vilela2015}.

\subsection{Direct Results}

Our two new metrics paint a complete picture of how this system will respond for both Bob and Eve, thus providing security analysis and system design constraints. The BE-CDF$^{bc}$ will show us the operating point for Bob to attain any desired level of reliability, and will also show us how Eve's decoding capability breaks down. The BER-CDF$^{ac}$ will then further enlighten us as to where we truly wish Eve to operate so as to guarantee (with probability essentially one) high BER at the output of her decoder. Coincidentally, this analysis also allows us to identify the jamming power advantage required during key transmission for the system to be successfully deployed~\cite{vilela2015}.

%\subsection{BE-CDF$^{bc}$}
%\todo{random interleaving for secrecy scheme \cite{vilela2015}? -- Must be cautious so as not to collide with the signal processing letters' paper}

% \begin{figure}
%  \centering
%    \begin{lpic}[]{Metric1_ourScheme(9cm,5.5cm)}
%    %\includegraphics[width=0.75\textwidth]{CrownePlazaQuad}
%    \begin{footnotesize}
%      \lbl[t]{2.5,71,90; $\color{blue}{\Pr}$}
%    \end{footnotesize}
%    \begin{scriptsize}
%      \lbl[t]{43,66.75; $\Pr$}
%    \end{scriptsize}
%    \end{lpic}
%  \caption{Metric 1.}
%  \label{fig:winner_pos_hist}
%\end{figure}

\begin{figure}
  \centering
   \ifCLASSOPTIONonecolumn
    \includegraphics[width=0.5\columnwidth]{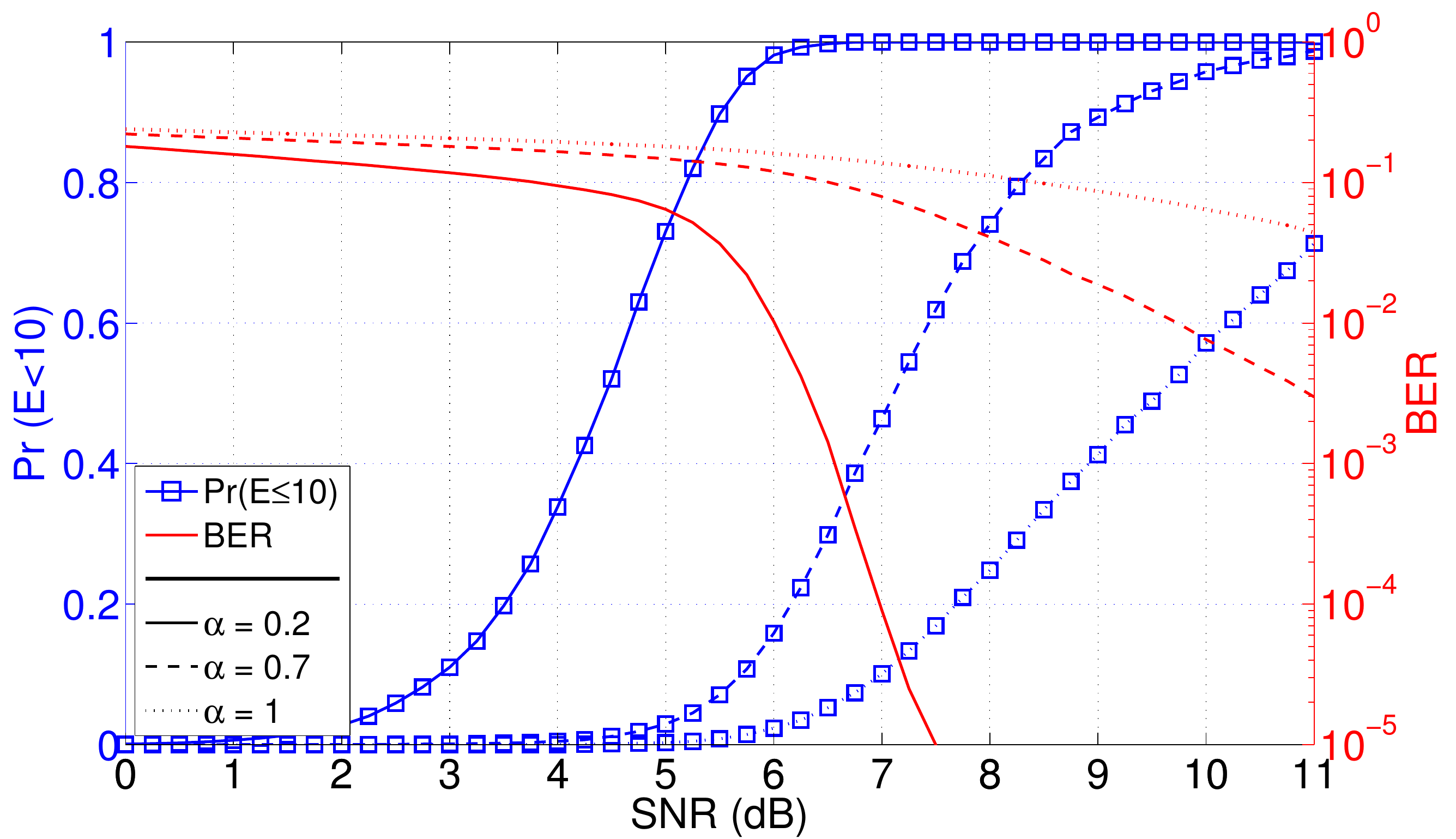}
   \else
    \includegraphics[width=\columnwidth]{Metric1_ourScheme_CDF_with_Marks.pdf}
   \fi
  \centering
  \caption{BE-CDF$^{bc}$ calculated when $t=10$ for three different effective jamming powers. These results anticipate the likelihood of decoder failure for Eve at $\alpha = 0.7$ for a BCH($127, 64$) code at around 0.9952 when Eve's Gaussian channel has SNR = 4 dB. If Bob experiences an effective $\alpha = 0.2$, then he can operate with BE-CDF$^{bc}=0.9975$ at 6.5 dB.}
  \label{fig:metric1_ourScheme}
\end{figure}

Let us assume that the effective jamming to Bob is $\alpha_B = 0.2$, while the effective jamming to Eve is $\alpha_E = 0.7$ (we also include $\alpha = 1$ in the figures for instructional purposes). 
The BE-CDF$^{bc}$ results apply to the BCH-encoded key bits and are given in Fig.~\ref{fig:metric1_ourScheme}, where we see that if Bob wishes to attain an overall BER around $10^{-3}$, the system must be designed to guarantee a BE-CDF$^{bc}$ value no lower than 0.9975. The interpretation of this value is that less than $1/4$ of 1\% of the transmitted key blocks should be decoded in error for Bob. Also according to Fig.~\ref{fig:metric1_ourScheme}, Bob achieves this performance if the SNR over his Gaussian channel is 6.5 dB or greater. We also note that the BE-CDF$^{bc}$ for Eve at an SNR of 4 dB is equal to 0.0048, meaning less than $1/2$ of 1\% of the time Eve will receive a key block for which she can correct all the errors if this BE-CDF$^{bc}$ value can be maintained.

\begin{figure}
  \centering
   \ifCLASSOPTIONonecolumn
    \includegraphics[width=0.5\columnwidth]{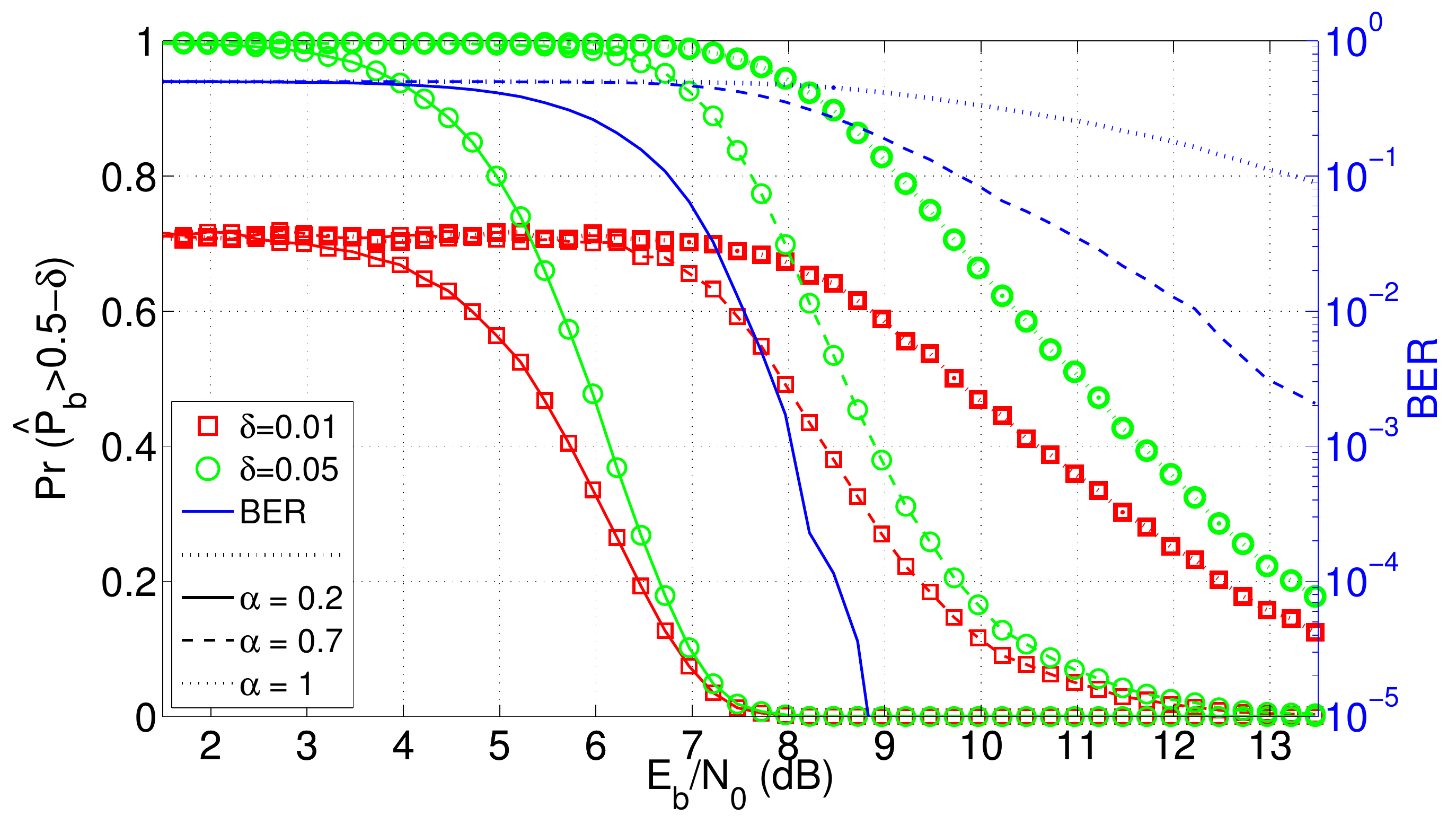}
   \else
    \includegraphics[width=\columnwidth]{Metric2_ourScheme.pdf}
   \fi
  \centering
  \caption{BER-CDF$^{ac}$ given for two delta values along with the BER. These three curves are given for three different effective jamming powers, and show that if Eve experiences jamming power $\alpha_E = 0.7$, then her BER over 753 message bits is guaranteed to be within $\delta = 0.05$ of 0.5 with high probability as long as her $E_b/N_0$ is no greater than 4.7 dB. This corresponds to an SNR value of approximately 4 dB.}
  \label{fig:metric2_ourScheme}
\end{figure}

To get the true feel for how Eve is affected by this scheme, however, we need to track the distribution of error proportion in Eve's guess of the message bits as a function of $E_b/N_0$ using the BER-CDF$^{ac}$ as depicted in Fig.~\ref{fig:metric2_ourScheme}. Here we see that for $\delta = 0.05$, we can attain $\Pr(\hat{P}_b>0.5-\delta) = 0.995$ at roughly $E_b/N_0 = 4.7$ dB, which corresponds to an SNR value of approximately 4 dB. These results indicate that for this scheme, insuring that Eve cannot correct all errors in the key is in fact sufficient for insuring a high proportion of errors in Eve's estimate of each short blocklength of message bits at the output of her decoder, which is exactly what we'd like to see in a practical physical-layer security scheme. For the sake of referring back to Fig.~\ref{fig:limitMetric2} for the limiting value of the BER-CDF$^{ac}$ metric, $\size_b$ for this scheme is the dimension of the LDPC code (880 bits) minus the blocklength of the BCH code (127 bits), because the BCH code only encodes the key bits and the remainder of the bits in the dimension of the LDPC code are dedicated to the message. This yields $\size_b = 753$ bits.

%\subsection{BER-CDF$^{ac}$}
%\todo{improving the puncturing and scrambling use cases?}

\subsection{Creating a Discrete Memoryless Channel}
Explicit secrecy code constructions exist that can provide information-theoretic security; however, only for discrete memoryless wiretap channels. As mentioned previously, currently known designs require either a noiseless main channel for legitimate communication or a degraded wiretap channel for the eavesdropper~\cite{Harrison2013}. Thus, we have two possible research directions for making these designs more practical to real end users. First, effort can be placed to design secrecy codes that operate over more realistic channels; and second, coding and/or signaling techniques may be leveraged to produce an \emph{effective} wiretap channel~\cite{Liang2011} over which we already know how to code for secrecy. In this section, we outline how our new metrics and the coding scheme explained in Section~\ref{sec:scheme} can be used to produce an effective discrete memoryless wiretap channel.

\begin{figure*}
  \centering
   \ifCLASSOPTIONonecolumn
    \includegraphics[width=\columnwidth]{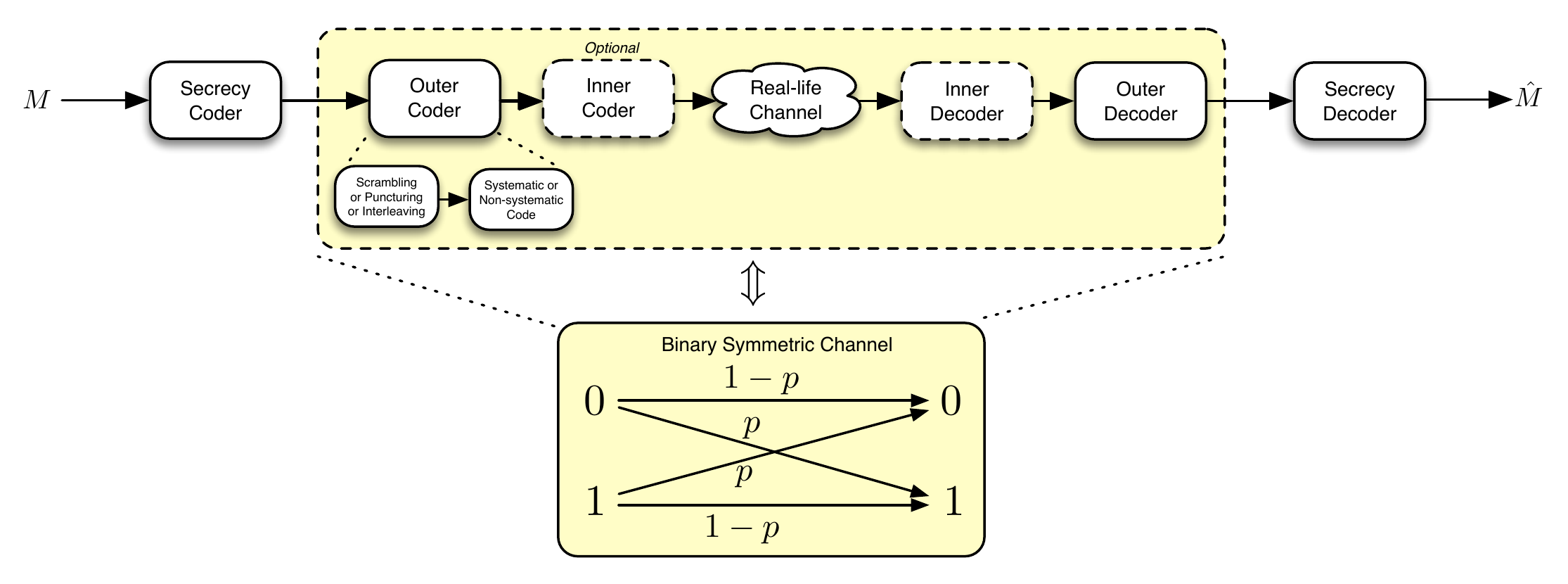}
   \else
    \includegraphics[width=2\columnwidth]{BSC_Fig2}
   \fi
  \centering
  \caption{A concatenated coding scheme may be utilized to provide an effective discrete memoryless wiretap channel, over which known explicit secrecy codes may operate for information-theoretic security.}
  \label{fig:BSC}
\end{figure*}

Consider again the results shown in Fig.~\ref{fig:metric2_ourScheme} that indicate an eavesdropper experiencing jamming power $\alpha_E = 0.7$ and $E_b/N_0 = 4.7$ dB over a Gaussian channel can expect error rates over 753-bit messages to have BER greater than 0.45 with probability very close to one. Since the analysis was conducted over short block lengths, we offer not just an average BER, but rather a low estimate of the BER over the channel. We now consider applying one more code on the outside of the entire scheme described in Section~\ref{sec:scheme}, as depicted in Fig.~\ref{fig:BSC}, and modeling the remaining blocks as an effective \ac{BSC}. The additional code added is one that can leverage this effective channel to bring about an information-theoretic security result (e.g.,~\cite{Mahdavifar2011}). 

In order to claim that the interior blocks in Fig.~\ref{fig:BSC} can truly be modeled as a \ac{BSC}, we need to verify three main properties of the \ac{BSC} in our system: (1) each bit should be erased independently from all other bits; (2) the probability $p$ of flipping each bit over the channel should be identical, and we need to identify its value; and (3) we need to ensure that soft information about the bit is either not available or impossible to use at the secrecy decoder. 

To ensure that bits within message blocks retain their independence of being in error, as required by the \ac{BSC} model, we need to apply an inter-block interleaver as the first subcode in the Outer Coder block in Fig.~\ref{fig:BSC} to spread information around as in~\cite{Harrison2011,Liang2011} and many other works. Although there may exist some correlations between flipped bits over the same transmitted packet, since all bits from every secrecy codeword are transmitted in different packets over the channel we effectively deliver independence between the bits at the secrecy codeword level, which is where we need independence for the secrecy code to work properly. 

In terms of identifying the probability $p$ that corresponds to the flipping of each bit over the channel, we'll use the lower bound given by BER-CDF$^{ac}$ as indicated above. By so doing, we provide an even stronger guarantee than identifying an average probability, since even short blocklengths maintain this probability of bit error with probability close to one. Bit error locations within secrecy codewords are kept uniformly random as a byproduct of the inter-codeword interleaving at the output of the secrecy encoder.

\begin{figure}
  \centering
   \ifCLASSOPTIONonecolumn
    \includegraphics[width=0.5\columnwidth]{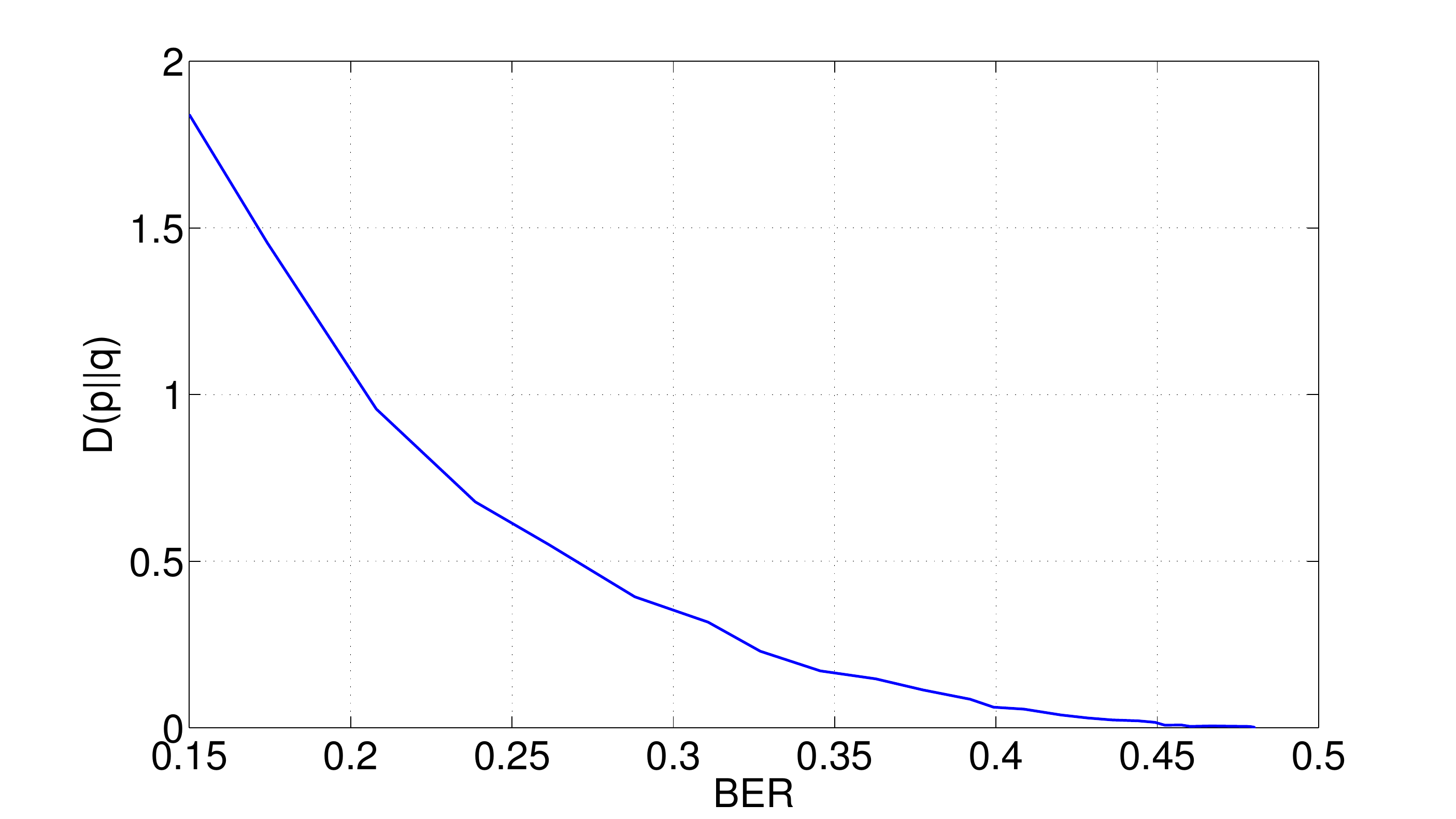}
   \else
    \includegraphics[width=\columnwidth]{KLdist2}
   \fi
  \centering
  \caption{Kullback-Leibler divergence between distributions of LLRs that correspond to bits in error and LLRs that correspond to correct bits at the output of an LDPC soft decoder, as a function of the hard-decision BER at the output of the decoder. As the BER approaches 0.5, the distributions become more alike, to the point where detecting a correct bit or a bit in error is impossible, even with soft information.}
  \label{fig:KLdist}
\end{figure}

Finally, we need to address this issue of soft information at the input of the secrecy decoder. Although soft information is technically available here, we must deduce whether or not the information is actually worth anything. In other words, what do log-likelihood ratios (LLRs) look like when the overall bit error rates at the output of an LDPC decoder are close to 0.5? LLRs can be approximated by Gaussian distributions with means centered at positive values if the bits should have a value of zero, and at negative values if the bits should have a value of one. The Gaussian approximation rule-of-thumb stems from the central limit theorem for likelihood ratios, where sums of random variables are calculated to give the ratio's next iteration~\cite{MoonCoding,Zheng2010}. The distribution of LLRs corresponding to bits in error is always symmetric and centered at zero since the decision threshold at the end of the soft iterative decoding algorithm is positioned directly between the distributions of LLRs corresponding to differing bit values. When the SNR is small enough that the code doesn't correct all the errors, distributions corresponding to bits in error and correct bits start to look very similar. In fact, when the noise completely overwhelms the coding scheme, each of these distributions tends to an approximate Gaussian distribution with mean zero and identical variances. It is this property that supplies an effective decoding \emph{threshold} for iteratively decodable codes~\cite{MoonCoding}. Finally, as the \ac{BER} approaches 0.5, the statistical difference between the distributions of LLRs for correct bits and bits in error becomes negligible. To demonstrate this, we show through simulation that the Kullback-Leibler (K-L) divergence~\cite{CoverAndThomas} between the two distributions approaches zero as the \ac{BER} approaches 0.5, where the K-L divergence is given as
\begin{equation}
  D(p || q) = \int\limits_x p(x) \log_2 \frac{p(x)}{q(x)},
\end{equation} 
and $p(x)$ represents the distribution of LLRs for correct bits while $q(x)$ represents the distribution of LLRs for bits in error at the output of a soft-information LDPC decoder. 
These results are given in Fig.~\ref{fig:KLdist}, where we observe $D(p||q)$ going to zero with increasing BER. Recognize that $D(p||q) = 0$ implies that there is no statistical difference between $p(x)$ and $q(x)$, or that the \emph{distance} between the two distributions is zero. It can be argued then, that as long as $D(p||q)$ is small enough, soft information at the output of an iterative decoder is unusable as it doesn't accurately depict any type of relationship between a bit's likelihood of being correct or in error.

The end result is that our new metrics mixed with the scheme from~\cite{vilela2015} can provide the effective channel model necessary for these information-theoretic designs to succeed. We see in~\cite{Harrison2013} that one type of secrecy code that may be able to offer secrecy over this channel is that given in~\cite{Mahdavifar2011}, where known advantageous (good for Bob, and bad for Eve) polarizations of bits in polar codes are used to transmit secret information over a symmetric eavesdropper's channel. This coding scheme is known to achieve strong secrecy at information rates approaching the secrecy capacity when the legitimate channel can be modeled as noiseless. For our case (where we've assumed that $\alpha_E = 0.7$, $\alpha_B = 0.2$, Bob's $\textrm{SNR} \geq 6.5$ dB, and Eve's $\textrm{SNR} \leq 4$ dB), supplying a probability of a flipped bit $p=0.45$ over an effective \ac{BSC} to an eavesdropper while maintaining an effectively noiseless main channel results in secrecy capacity $C_s = C_m - C_w = p$ bits per channel use, where $C_m$ and $C_w$ signify the channel capacities of the main and wiretap channel, respectively~\cite{wyner:75,csiszar:78,CoverAndThomas}. 

The approach outlined here, where we manufacture a wiretap channel over which additional secrecy codes can be utilized, can be extended to produce other effective discrete memoryless wiretap channels as well that may form ideal backdrops for other code designs to operate in more realistic environments.

\section{Conclusions}
\label{sec:conclusion}

In this paper we have discussed the landscape of physical-layer security coding metrics. We note that most measures in use today rely on information-theoretic analysis as blocklengths tend to infinity, or use mean BER, both of which give asymptotic results that have limited meaning for short blocklength codes. We have proposed two new metrics that effectively employ CDFs to provide a lower bound on the security levels based on BER. Such an approach provides a stronger guarantee of secrecy over realistic channel models than simply using mean BER to estimate performance, and yet our metrics retain their simplicity of calculation making them directly adaptable to real-world communication systems. We have also shown how these new metrics may be used to reduce realistic channel model environments to simpler models over which known secrecy codes may be implemented to achieve information-theoretic security.

%Error rate associated with an error cdf with (1-$\epsilon$).

% or


%% \appendices
%% \section{Proof of the First Zonklar Equation}
%% Appendix one text goes here.

%% % you can choose not to have a title for an appendix
%% % if you want by leaving the argument blank
%% Appendix two text goes here.

%% \section*{Acknowledgment}

%% The authors would like to thank...

\ifCLASSOPTIONcaptionsoff
  \newpage
\fi



%\bibliographystyle{IEEEtran}
\bibliographystyle{IEEEtran}
\bibliography{IEEEabrv,refs}

%% \begin{thebibliography}{1}

%% \bibitem{IEEEhowto:kopka}
%% H.~Kopka and P.~W. Daly, \emph{A Guide to \LaTeX}, 3rd~ed.\hskip 1em plus
%%   0.5em minus 0.4em\relax Harlow, England: Addison-Wesley, 1999.

%% \end{thebibliography}


%% \begin{IEEEbiography}{Michael Shell}
%% Biography text here.
%% \end{IEEEbiography}

%% % if you will not have a photo at all:
%% \begin{IEEEbiographynophoto}{John Doe}
%% Biography text here.
%% \end{IEEEbiographynophoto}

%\newpage

%% \begin{IEEEbiographynophoto}{Jane Doe}
%% Biography text here.
%% \end{IEEEbiographynophoto}


%\vfill


\end{document}